\documentclass[aps,preprintnumbers,amsmath]{revtex4}
\usepackage{amssymb}
\usepackage{graphicx}
\usepackage{dcolumn}
\usepackage{bm}

\begin{document}

\title{Collective and fractal properties of pion jets \\ in the
four-velocity space at intermediate energies}

\author{V.A. Okorokov} \email{VAOkorokov@mephi.ru; okorokov@bnl.gov}
\affiliation{National Research Nuclear University "MEPhI",
Kashirskoe Shosse 31, 115409 Moscow, Russia}

\author{A.K. Ponosov}
\affiliation{National Research Nuclear University "MEPhI",
Kashirskoe Shosse 31, 115409 Moscow, Russia}

\author{F.M. Sergeev}
\affiliation{National Research Nuclear University "MEPhI",
Kashirskoe Shosse 31, 115409 Moscow, Russia}

\date{\today}

\begin{abstract}
Experimental results are presented for study of collective and
fractal properties of soft pion jets in the space of relative
four-dimensional velocities. Significant decreasing is obtained
for mean square of second particle distances from jet axis for
pion-proton interactions at initial energies $\sim 3$ GeV in
comparison with hadron-nuclear collisions at close energies. The
decreasing results in power dependence of distance variable on
collision energy for range $\sim 2 - 4$ GeV. The observation
allows us to estimate the low boundary of manifestation of color
degree of freedom in pion jet production. Cluster dimension values
were deduced for pion jets in various reactions. Fractional values
of this dimension indicate on the manifestation of fractal-like
properties by pion jets. Changing of mean kinetic energy of jet
particles and fractal dimension with initial energy increasing is
consistent with suggestion for presence of color degrees of
freedom in pion jet production at intermediate energies.

\textbf{PACS}
13.75.-n
\end{abstract}

\maketitle

\section{\label{intro}Introduction}
Investigation of the mixed phase, transition from meson-nucleon to
quark-gluon degrees of freedom is now one of the most actual and
important tasks of the world program of researches in the field of
strong interactions. Experimental study of hadron-hadron and
hadron-nuclear interactions can give the important information
concerning manifestation of new (color) degrees of freedom. In
various fields of physics the beginning of manifestation of new
degrees of freedom and transition processes are accompanied by
presence of self-affine, fractal features for collective effects.
Thus, study of collective and geometrical (fractal) properties, in
particular, soft hadron jets at intermediate energies can give the
new important information concerning hadronization mechanisms,
non-perturbative physics and transition to manifestation of color
degrees of freedom in collective phenomena.

A relativistic invariant method for investigation of collective
effects in multiparticle production processes of the type
$\mbox{I} + \mbox{II} \to 1+2 +...$ was proposed in
\cite{Baldin-DoklANUSSR-222-1064-1975}. Features of this method
for events with two jets were in detail considered in
\cite{Kiselevich-YaF-57-225-1994,Mikhailichenko-YaF-62-1787-1999}.
In this case the secondary particles emitted in the regions of
target and beam fragmentation can be separated by means of a
relativistic invariant variable $X_{\footnotesize\mbox{l}} =0.1 -
0.2$. One of the main variables of the method is
$b_{k}=-\left(V-U_{k}\right)^{2}$ -- a square of distance $k$-th
particle from an axis of jet $V$ in the space of four-dimensional
velocities $U_{i}=P_{i}/m_{i},~i = \mbox{I, II},1,2... $
\cite{Baldin-DoklANUSSR-222-1064-1975}. It should be stressed that
in the framework of this approach the range of values $10^{-2}
\leq b_{k} \sim 1$ corresponds to transition from domination of
meson-nucleon degrees of freedom to manifestation of internal
structure of initial particles and, thus, to quark-gluon degrees
of freedom in processes of production of secondary particles.

In \cite{Okorokov-NSMEPhI-218-2000} it was suggested to study the
geometrical properties of pion jets in four-velocity space by
means of cluster dimension, $D$, defined by a relation between
number of particles in jet, $N\left(b_{k}\right)$, and its radius
$N\left(b_ {k}\right) \propto b_{k}^{D/2}$.

In this paper, collective and geometrical properties of soft pion
jets have been studied using the space of relative
four-dimensional velocities for following reactions:
\begin{equation}
\pi^{-}+\mbox{p} \to \mbox{N}+k\pi^{-}+l\pi^{+}+j\pi^{0},~~k=2,
l=1,2, j \geq 0,~~P_{0}=(3.93 \pm 0.01)~\mbox{GeV/c};
\label{eq:reac-1}
\end{equation}
\vspace*{-0.7cm}
\begin{equation}
\pi^{+}+\mbox{p} \to
\begin{cases}
\mbox{N}+K^{+}\left(K^{0}\right)+\bar{K}^{0}\left(K^{-}\right)+k\pi^{-}+l\pi^{+}+j\pi^{0},&
P_{0}=(3.90 \pm 0.30)~\mbox{GeV/c}\\
\mbox{B}_{s}+K^{0,+}+k\pi^{-}+l\pi^{+}+j\pi^{0},&k,j=0,1, l \geq
0; \end{cases} \label{eq:reac-2}
\end{equation}
\vspace*{-0.5cm}
\begin{equation}
\pi^{+}+\mbox{p} \to \mbox{N}+k\pi^{-}+l\pi^{+}+j\pi^{0},~~k=1,2,
l=2-4, j=0,1,~~P_{0}=(4.23 \pm 0.08)~\mbox{GeV/c};
\label{eq:reac-3}
\end{equation}
\vspace*{-0.5cm}
\begin{equation}
\pi^{-}+\left(\mbox{C}_{2}\mbox{F}_{5}\mbox{Cl}_{3}\right) \to
\Lambda^{0}+K^{0}+k\pi^{-}+l\pi^{+}+m\mbox{p}+X,~~k,l,m
> 0,~~P_{0}=(3.86 \pm 0.04)~\mbox{GeV/c}; \label{eq:reac-4}
\end{equation}
\vspace*{-0.5cm}
\begin{equation}
\pi^{-}+\mbox{Ne} \to k\pi^{-}+l\pi^{+}+m\mbox{p}+X,~~k,l,m
> 0,~~P_{0}=(6.20 \pm 0.10)~\mbox{GeV/c}; \label{eq:reac-5}
\end{equation}
Here $P_{0}$ -- beam momentum, $\mbox{N=p,n}$,
$\mbox{B}_{s}=\Lambda^{0}, \Sigma^{0,\pm}$ are final state strange
baryon, moreover $l=1,2$ at $\mbox{B}_{s}=\Lambda^{0}$ and $l=0-3$
at $\mbox{B}_{s}=\Sigma^{0,\pm}$ in the reaction (\ref
{eq:reac-2}). The technique of experiments and a selection
criteria are described in detail in
\cite{Mikhailichenko-YaF-62-1787-1999,Chaloupka-PLB-44-1973-211}.
The general statistics for reactions (\ref {eq:reac-1}) -- (\ref
{eq:reac-5}) is more than $1.9 \cdot 10^{5}$ events
\cite{Okorokov-IntSemHEP-154-2006}.

\section{\label{sec2}Collective properties of pion jets}
\begin{figure}[t!]
\includegraphics[width=17.5cm]{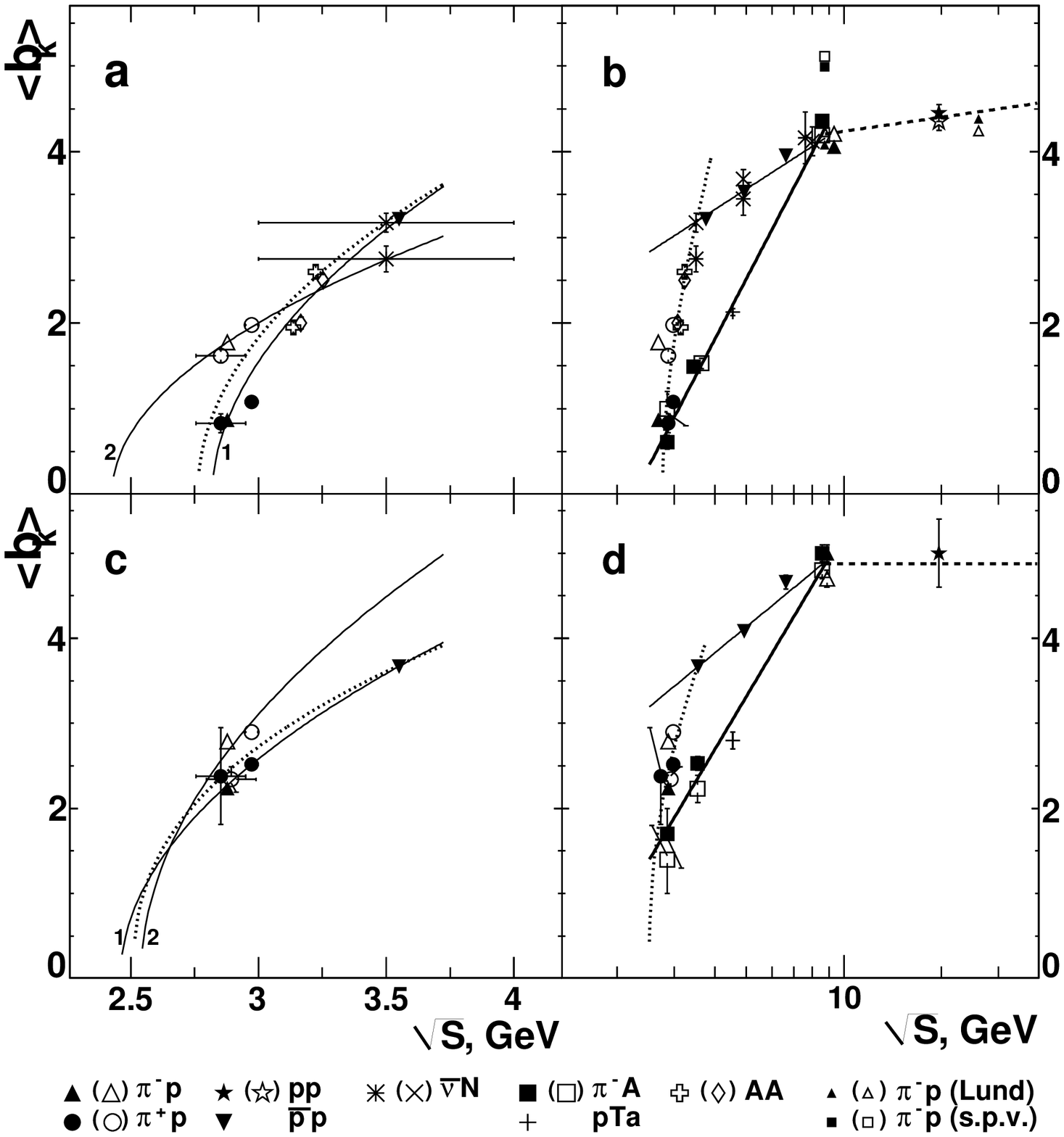}
\caption{Dependence of
$\langle b_{k}^{\pi}\rangle$ on $\sqrt{s}(W)$ at $X_{\textstyle
\mbox{l}}=0.1$ (a,b) and $X_{\textstyle \mbox{l}}=0.2$ (c,d).
Experimental data for reactions (\ref{eq:reac-3}) --
(\ref{eq:reac-5}) are from
\cite{Kiselevich-YaF-57-225-1994,Mikhailichenko-YaF-62-1787-1999,Okorokov-IntSemHEP-154-2006},
for $\bar{\mbox{p}}\mbox{p}$ at 5.7, 12 and 22.4 GeV/c,
$\bar{\nu}\mbox{N}$ at W=3-4, 4-6 and $\geq 6$ GeV, $\mbox{pp}$ at
205 GeV/c, $\pi^{-}\mbox{p}$ at 40 GeV/c,
$\mbox{pTa}+\mbox{p}\left(\mbox{C}_{3}\mbox{H}_{8}\right)$ at 10
GeV/c, $\pi^{-}\mbox{C}$ at 40 GeV/c, $\mbox{CC}$ at 4.2 GeV/c/A,
$\mbox{MgMg}$ at 4.5 GeV/c/A, model calculations -- from
\cite{JINRRapidCom-N16-86-24-1986}. The curves are described in
the main body of the text.}\label{fig:1-NegPions}
\end{figure}

Fig.\ref {fig:1-NegPions} shows dependencies of mean square of
distance from axis of pion jet in the four-velocity space --
$\langle b_{k}\rangle$ -- on initial energy $\sqrt{s}$ (mean total
energy of final hadron state in the c.m. frame, $W$, for
$\bar{\nu}\mbox{N}$) for secondary $\pi^{-}$ mesons at cutoff
$X_{\footnotesize\mbox{l}} =0.1$ (a,b) and 0.2 (c,d) for various
interactions.

At initial energies $\sqrt{s} > 8$ the dependence $\langle
b_{k}\rangle \left(\sqrt{s}\right)$ has been fitted by logarithmic
function
\begin {equation}
\langle b_{\textstyle k}\rangle=a_{1} +a_{2}
\ln\left(s/s_{0}\right), \label {eq:6}
\end {equation}
where $s_{0}=1$ GeV$^{2}$. Results of above fit are presented on
Fig.\ref{fig:1-NegPions}b,d (dashed line) and in Table
\ref{table:1}. Only experimental data are fitted at
$X_{\footnotesize\mbox{l}}=0.1$ for energy range under considered.
As seen the Lund model calculations for $\pi^{-}\mbox{p}$
interactions at 40 GeV/c are in good agreement with experimental
data, unlike results obtained by the model of homogeneously filled
phase volume (Fig.\ref {fig:1-NegPions}b). The fit with taking
into account results of Lund model calculations for
$\pi^{-}\mbox{p}$ at 40 and 360 GeV/c gives parameter values which
coincide within errors with the values indicated in Table
\ref{table:1} at some improvement of fit quality
($\chi^{2}$/n.d.f.=4.53). The value of $a_{2}$ parameter in (\ref
{eq:6}) coincides with zero within errors at
$X_{\footnotesize\mbox{l}}=0.2$ in the energy domain $\sqrt{s} >
8$ GeV. Therefore data were fitted at the fixed value $a_{2}=0$,
that allows to obtain better quality of approximation. However, it
is impossible to exclude unambiguously the weak logarithmic growth
for $\langle b_{k}\rangle$ in accordance with (\ref {eq:6}) taking
into account insignificant ensemble of accessible experimental
data. Earlier in \cite{Mikhailichenko-YaF-62-1787-1999}
statistically reasonable description by function (\ref {eq:6}) had
been obtained for experimental data for hadron-hadron and
$\bar{\nu}\mbox{N}$ reactions, and also for hadron-nuclear
interactions, at $3.5 < \sqrt{s} <9$ GeV (solid lines on Fig.\ref
{fig:1-NegPions}b,d). Thus, dependence $\langle b_{k}\rangle
\left(\sqrt{s}\right)$ supposes universal approximation
(\ref{eq:6}) at $\sqrt {s} > 3.5$ GeV for a wide class of
interactions at any values $X_{\footnotesize\mbox{l}}$ under
study.

The hypothesis has been suggested in
\cite{Mikhailichenko-YaF-62-1787-1999} about change of dynamic
regime at $\sqrt{s} < 3 - 4$ GeV, reflected in the behaviour of
underlying dependence. This energy domain has been investigated in
more details in the present paper. Dependencies $\langle
b_{k}\rangle \left(\sqrt{s}\right)$ are shown on
Fig.\ref{fig:1-NegPions}a,b in collision energy range $\sqrt{s} <
4$ GeV with taking into account of uncertainties of initial
momenta for various types of interactions with exception of
hadron-nuclear ones at $X_{\footnotesize\mbox{l}}=0.1$ and 0.2,
respectively. The results obtained here for a pion-proton
reactions (\ref{eq:reac-1}) and (\ref {eq:reac-2}) indicate
clearly on the change of behaviour of dependence $\langle
b_{k}\rangle$ at $\sqrt{s} \sim 3$ GeV (Fig.\ref{fig:1-NegPions}),
confirming the our earlier suggestion
\cite{Mikhailichenko-YaF-62-1787-1999}. Results for symmetric
nuclear collisions agree well with the general tendency
(Fig.\ref{fig:1-NegPions}a,b) at $X_{\footnotesize\mbox{l}}=0.1$.
Dependence of $\langle b_{k}\rangle$ values on $\sqrt{s}(W)$ has
been fitted by following power function at $\sqrt{s} < 4$ GeV:
\begin {equation}
\langle b_{\textstyle k}\rangle=a_{1}
\left(\sqrt{s/s_{0}}-a_{2}\right)^{a_{3}},~~~\sqrt{s/s_{0}} \geq
a_{2}. \label {eq:7}
\end {equation}
The results of the above fit are shown on Fig.\ref{fig:1-NegPions}
(dotted lines) and in Table \ref{table:1}. In
\cite{Mikhailichenko-YaF-62-1787-1999,Okorokov-IntSemHEP-154-2006}
it has been shown, that properties of soft pion jets depend on
region of fragmentation in the energy domain under study. This
effect leads to significant distinction of corresponding $\langle
b_{k}\rangle$ values. Therefore sharp behaviour of dependence
$\langle b_{k}\rangle \left(\sqrt{s}\right)$ and disorder of
experimental points result in statistically unacceptable fit
quality. However, as seen from Fig.\ref{fig:1-NegPions}, power
function agrees well enough with experimental data at all values
of $X_{\footnotesize\mbox{l}}$ at qualitative level.

Experimental data samples were fitted for target fragmentation
region (curves 1 on Fig.\ref{fig:1-NegPions}a,c) and for beam
fragmentation one (curves 2 on Fig.\ref{fig:1-NegPions}a,c)
separately. The values of fit parameters are listed in Table
\ref{table:2}. Substantial improvement of fit quality is observed
for any fragmentation region and values of
$X_{\footnotesize\mbox{l}}$. The data samples at
$X_{\footnotesize\mbox{l}}=0.2$ were fitted at the fixed value of
parameter $a_{3}=0.5$ taking into account the results obtained at
soft cutoff and volumes of accessible data samples at
$X_{\footnotesize\mbox{l}}=0.2$. Thus, dependence $\langle
b_{k}\rangle \left(\sqrt{s}\right)$ is described by power function
(\ref {eq:7}) at $\sqrt {s} < 4$ GeV for studied types of
interactions ($\mbox{hh}, \bar{\nu}\mbox{N}, \mbox{AA}$) both for
separate samples for various fragmentation regions and for total
data ensemble. It is important to note, that in general the power
behaviour is one of the characteristic features of transition
domain in which new degrees of freedom of investigated system
begin to be manifest. Value of parameter $a_{2}$ can be put in
correspondence with the energy (in GeV), $\sqrt{s_{c}}$, at which
the internal structure of interacting particles, i.e. new degrees
of freedom, begins to manifest in the pion jet production. Thus,
the estimation is obtained, that new (color) degrees of freedom
begin to be manifest experimentally at $\sqrt{s_{c}} \simeq 2.5 -
2.8$ GeV in jet production of pions.

\begin{figure}[t!]
\includegraphics[width=17.5cm]{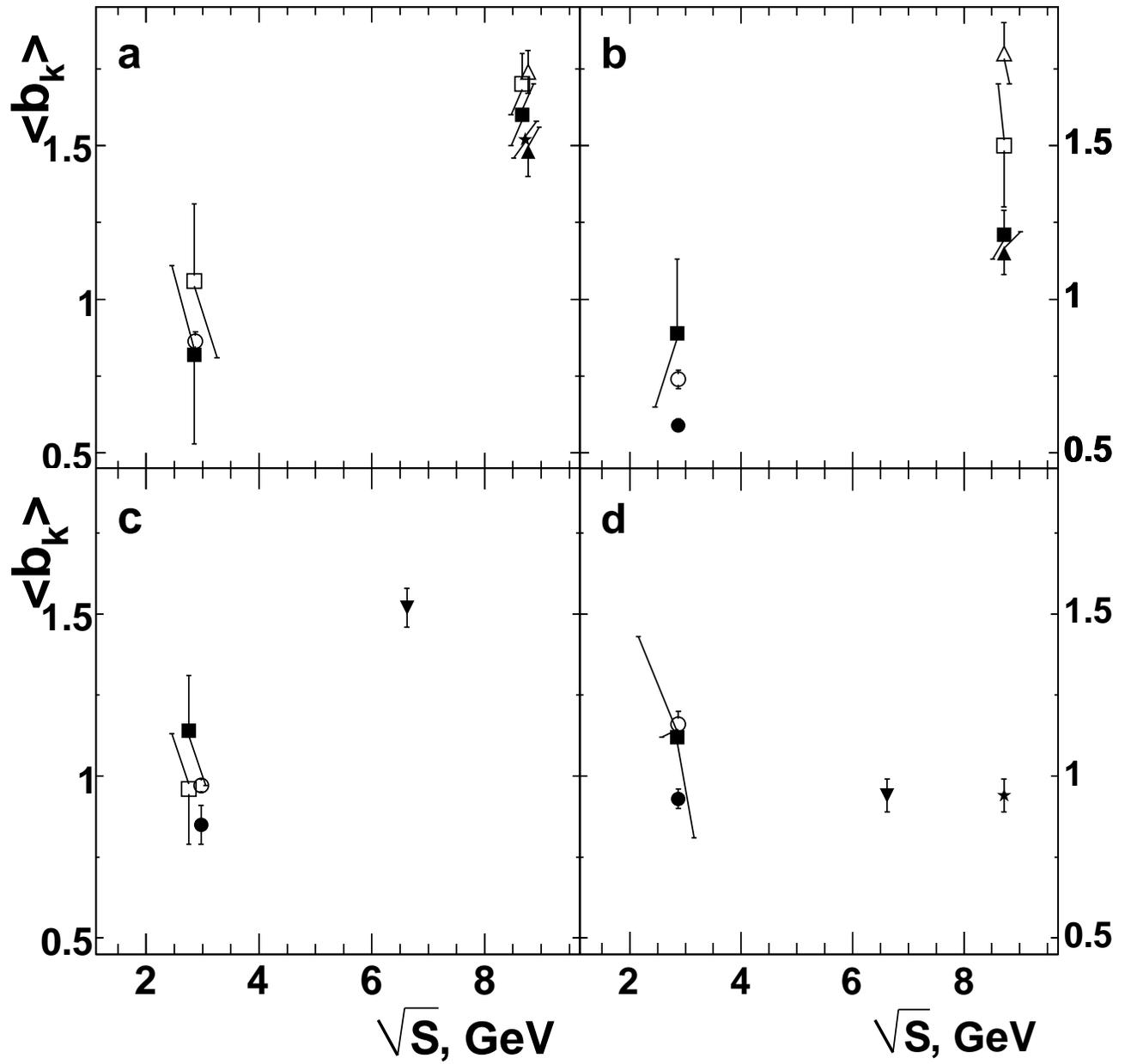}
\caption{Values of $\langle b_{k}\rangle$ versus collision energy
for strange particles at $X_{\footnotesize \mbox{l}}=0.1$ (a,b)
and $X_{\footnotesize \mbox{l}}=0.2$ (c,d). Left column
corresponds to $K^{0}$-mesons (a,c), right column -- to
$\Lambda^{0}$-hyperons (b,d). Experimental points for target
(beam) fragmentation region are marked as follows:
{\large$\bullet$} ({\large$\circ$}) -- reaction (\ref{eq:reac-2}),
$\blacktriangledown$ -- $\bar{\mbox{p}}\mbox{p}$ at 22.4 GeV/c
\cite{Baldin-PrJINR-820-1985,JINRRapidCom-N16-86-24-1986}
{\large$\star$}
 -- $\mbox{pp}$ at 40 GeV/c \cite{Baldin-PrJINR-820-1985},
 $\blacktriangle$ ($\vartriangle$) --
$\pi^{-}\mbox{p}$ at 40 GeV/c \cite{Baldin-PrJINR-820-1985},
$\blacksquare$ ($\square$) --
$\pi^{-}\left(\mbox{C}_{2}\mbox{F}_{5}\mbox{Cl}_{3}\right),~\pi^{-}\mbox{C}$
at 3.9 and 40 GeV/c \cite{Baldin-PrJINR-820-1985},
respectively.}\label{fig:2-StrangeParticle}
\end{figure}

Dependencies $\langle b_{k}\rangle \left(\sqrt{s}\right)$ are
presented at Fig.\ref{fig:2-StrangeParticle} for secondary neutral
strange particles, $K^{0}$ mesons (a,c) and $\Lambda^{0}$ hyperons
(b,d), at various values of $X_{\footnotesize\mbox{l}}$. As seen,
usually, $\langle b_{k}\rangle$ value is smaller for target
fragmentation region, than corresponding $\langle b_{k}\rangle$
for beam fragmentation one. The reactions (\ref {eq:reac-2}) and
(\ref {eq:reac-4}) investigated here with absence of strange
particles in an initial state and with its production in a final
state can be considered as additional argument in favour of
manifestation of quark degrees of freedom at $\langle b_{k}\rangle
\sim 1$ at intermediate energies. Accessible samples of results is
much less in case of neutral strange particles, than for $\pi^{-}$
mesons. Therefore it is possible to study qualitative behaviour of
$\langle b_{k}\rangle \left(\sqrt{s}\right)$ at
Fig.\ref{fig:2-StrangeParticle} only. For $K^{0}$ mesons as well
as for secondary pions, the increasing of $\langle b_{k} \rangle$
at growth of initial energy is observed at any
$X_{\footnotesize\mbox{l}}$ under considered
(Fig.\ref{fig:2-StrangeParticle}a,c). One can see that behaviour
of $\langle b_{k}\rangle \left(\sqrt{s}\right)$ for $\Lambda^{0}$
hyperons at $X_{\footnotesize\mbox{l}}=0.1$
(Fig.\ref{fig:2-StrangeParticle}b) is similar to corresponding
dependence for $K^{0} $ mesons (Fig.\ref{fig:2-StrangeParticle}a),
however $\langle b_{k}\rangle$ values for $\Lambda^{0}$ do not
depend on collision energy practically at more hard cutoff
$X_{\footnotesize\mbox{l}}=0.2$
(Fig.\ref{fig:2-StrangeParticle}d).
\begin{table}
\begin{minipage}[b]{.96\linewidth}
\begin{minipage}[t]{.46\linewidth}
\caption{Fit parameters for dependence $\langle b_{\textstyle
k}\rangle\left(\sqrt{s}\right)$.} \label{table:1}
\bigskip
\setlength{\extrarowheight}{2pt}
\begin{tabular}{|l|c|c|p{2.0cm}|} \hline
\multicolumn{1}{|c|}{Fit parameter}
&\multicolumn{1}{|c|}{$X_{\footnotesize \mbox{l}}=0.1$}
&\multicolumn{1}{|c|}{$X_{\footnotesize \mbox{l}}=0.2$} \\
\hline \multicolumn{3}{|c|}{$\mbox{hh}-,~\mbox{hA}-$interactions
($\sqrt{s} > 8$ GeV)} \\
\hline
$a_{1}$        &$3.7 \pm 0.2$   & $4.88 \pm 0.05$  \\
$a_{2}$        &$0.12 \pm 0.05$   & $0.0$ (fixed) \\
$\chi^{2}$/n.d.f. & 7.20             & 1.71 \\
\hline \multicolumn{3}{|c|}{$\mbox{hh}-,~\mbox{AA}-$interactions
($\sqrt{s} < 4$ GeV)} \\
\hline
$a_{1}$           &$3.69 \pm 0.02$ & $3.63 \pm 0.05$  \\
$a_{2}$           &$2.76 \pm 0.01$ & $2.51 \pm 0.03$  \\
$a_{3}$           &$0.49 \pm 0.01$ & $0.40 \pm 0.01$  \\
$\chi^{2}$/n.d.f. & 133             & 9.25  \\[1mm]
\hline
\end{tabular}
\end{minipage}\hfill
\begin{minipage}[t]{.50\linewidth}
\caption{Parameter values for $\langle b_{\textstyle
k}\rangle\left(\sqrt{s}\right)$ at $\sqrt{s} < 4$ GeV. \\
Separated fitting for target and beam fragmentation regions.}
\label{table:2}
\bigskip
\setlength{\extrarowheight}{2pt}
\begin{tabular}{|l|c|c|c|c|p{2.0cm}|} \hline
\multicolumn{1}{|c|}{$X_{\footnotesize \mbox{l}}$}
&\multicolumn{4}{|c|}{Fit parameters} \\ \cline{2-5}
 & $a_{1}$ & $a_{2}$ & $a_{3}$ & $\chi^{2}$/n.d.f. \\
\hline \multicolumn{5}{|c|}{target fragmentation}   \\
\hline
0.1 &$3.79 \pm 0.03$ & $2.82 \pm 0.02$ & $0.52 \pm 0.05$ & 5.70  \\
0.2 &$3.52 \pm 0.07$ & $2.46 \pm 0.04$ & $0.5$ (fixed)   & 0.09 \\
\hline \multicolumn{5}{|c|}{beam fragmentation} \\
\hline
0.1 &$2.65 \pm 0.13$ & $2.43 \pm 0.04$ & $0.50 \pm 0.35$ & 5.19  \\
0.2 &$4.6 \pm 1.9$   & $2.5 \pm 0.3$   & $0.5$ (fixed)   & 2.29\\
[1mm] \hline
\end{tabular}
\end{minipage}
\end{minipage}
\end{table}

The dependence of $\langle b_{k}\rangle$ value on particle mass
has been investigated for secondary
$\pi^{0,\pm},~K^{0,\pm},~\Lambda^{0}$ and $\Sigma^{0,\pm}$
particles from reaction (\ref {eq:reac-2}). The relativistic
invariant variable $X_{\footnotesize \mbox {l}}$ was used for
separation of various fragmentation region for all particle types
with exception of $\Sigma$ hyperons. The Feynman variable
$x_{\footnotesize\mbox{F}}$ was used for separation of more heavy
$\Sigma$ particles on fragmentation region. Significant decreasing
of mean square of distance from jet axis is observed for kaons in
comparison with pions with the weaker decreasing at further
increasing of particle mass that agrees with the behaviour of
similar dependence at higher energies $\sqrt{s} \simeq 8.7$ GeV
qualitatively \cite{Okorokov-IntSemHEP-154-2006}.

\begin{figure}[t!]
\includegraphics[width=17.5cm]{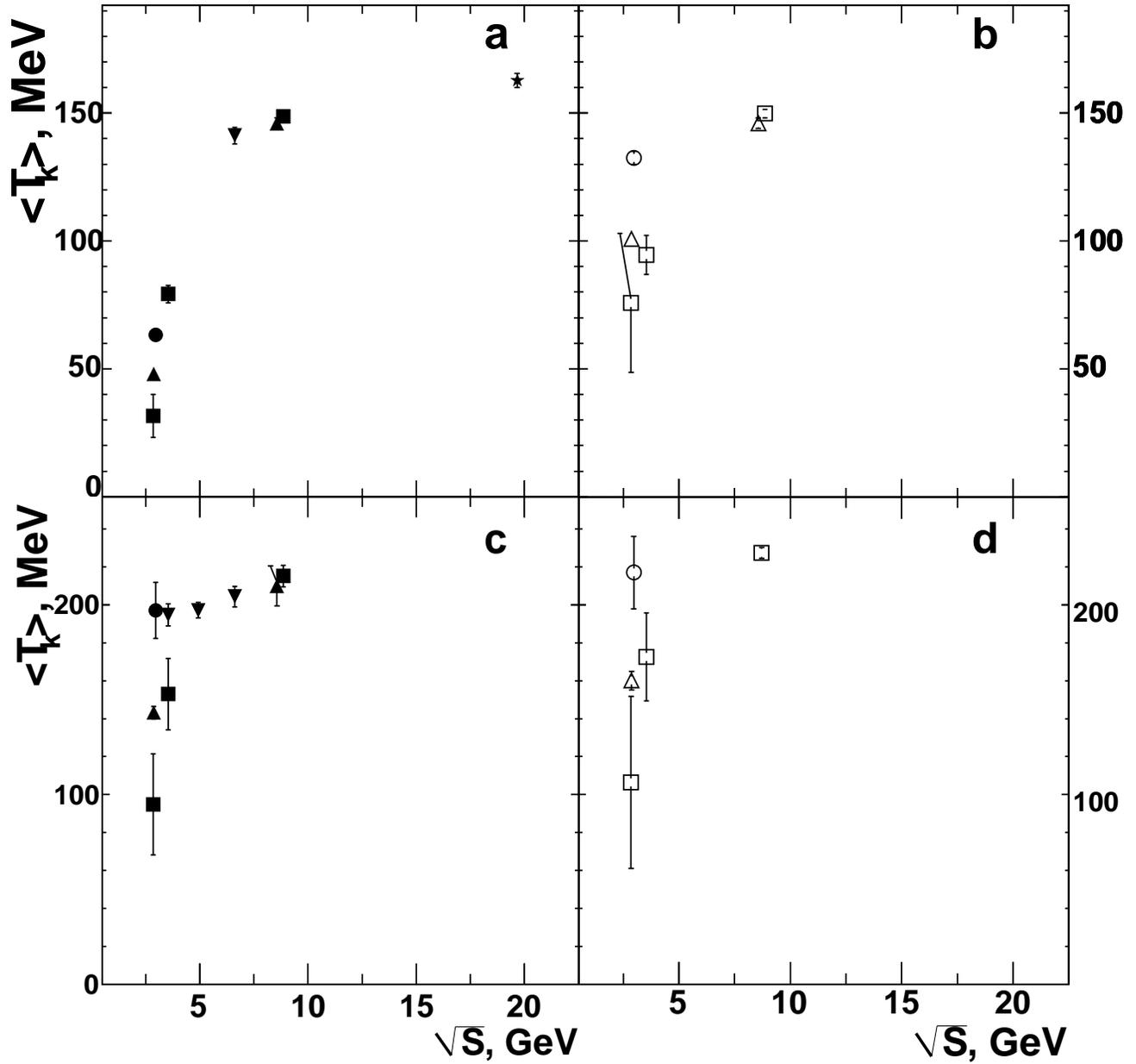}
\caption{Dependence $\langle
T_{k}\rangle \left(\sqrt{s}\right)$ for $\pi^{-}$-mesons at
$X_{\footnotesize \mbox{l}}=0.1$ (a,b) and $X_{\footnotesize
\mbox{l}}=0.2$ (c,d). Left column corresponds to the target
fragmentation region (a,c), right column -- to the beam
fragmentation (b,d). Experimental points for target (beam)
fragmentation region are marked as follows: {\large$\bullet$}
({\large$\circ$}) -- $\pi^{+}\mbox{p}$ at 4.2 GeV/c,
$\blacktriangledown$ -- $\bar{\mbox{p}}\mbox{p}$ at 5.7, 12 and
22.4 GeV/c, {\large$\star$} -- $\mbox{pp}$ at 205 GeV/c,
$\blacktriangle$ ($\vartriangle$) -- $\pi^{-}\mbox{p}$ at 3.9 and
40 GeV/c, $\blacksquare$ ($\square$) --
$\pi^{-}\left(\mbox{C}_{2}\mbox{F}_{5}\mbox{Cl}_{3}\right),~\pi^{-}\mbox{Ne},~\pi^{-}\mbox{C}$
at 3.9, 6.2 and 40 GeV/c,
respectively.}\label{fig:3-Tkin-vs-Energy}
\end{figure}

The important characteristic of applied approach is the mean
kinetic energy of particles in jet in the its rest frame (this
parameter is called "temperature" often), $ \langle T_{k}\rangle$,
which is calculated on the basis of fitting of invariant
$F\left(b_{k}\right)$-distributions
\cite{JINRRapidCom-N16-86-24-1986}. Fig.\ref{fig:3-Tkin-vs-Energy}
shows dependencies $\langle T_{k}\rangle \left(\sqrt{s}\right)$
for pion jets in the target fragmentation region (a,c) and for
beam fragmentation one (b,d). The published distributions $1/N
dN/db_{k}$ \cite{Baldin-PrJINR-142-1987} have been used for an
estimation of $\langle T_{k}\rangle$ in case of
$\bar{\mbox{p}}\mbox{p}, \mbox{pp}, \pi^{-}\mbox{C}$ interactions,
results for $\pi^{-}\mbox{p}$ reaction at 40 GeV/c are taken from
\cite{Baldin-ECHAYA-29-577-1998}. Accessible experimental data
allow to study behaviour of $\langle T_{k}\rangle
\left(\sqrt{s}\right)$ at a qualitative level. The increasing of
$\langle T_{k} \rangle$ is observed with growth of initial energy
both for hadron-hadron and for hadron-nuclear collisions. It is
important to note significant increasing of $\langle T_{k}\rangle$
in a narrow range of $\sqrt{s}$ for reactions (\ref {eq:reac-1}),
(\ref {eq:reac-3}), and much smoother growth at the further
increasing of initial energy. Additional experimental
investigations are necessary for interactions of various types at
$\sqrt{s} \sim 3 - 20$ GeV, however, the obtained results do not
contradict the hypothesis above concerning manifestation of new
degrees of freedom in processes of soft pion jet production at
$\sqrt{s} \sim 3$ GeV.

The results obtained for $\pi^{-}$ mesons from jets (clusters) of
secondary particles of various types in proton-nuclear and in
nuclear-nuclear interactions with using of normalized relative
four-velocity space for central region and for beam fragmentation
\cite{Angelov-RapidJINR-2-4-1990} are agreed with behaviour of $
\langle T_{k}\rangle \left(\sqrt{s}\right)$ at
$X_{\footnotesize\mbox {l}}=0.1$ for beam fragmentation
(Fig.\ref{fig:3-Tkin-vs-Energy}b) qualitatively.

\section{\label{sec3}Cluster dimensions for jets}
\begin{figure}[t!]
\includegraphics[width=17.5cm]{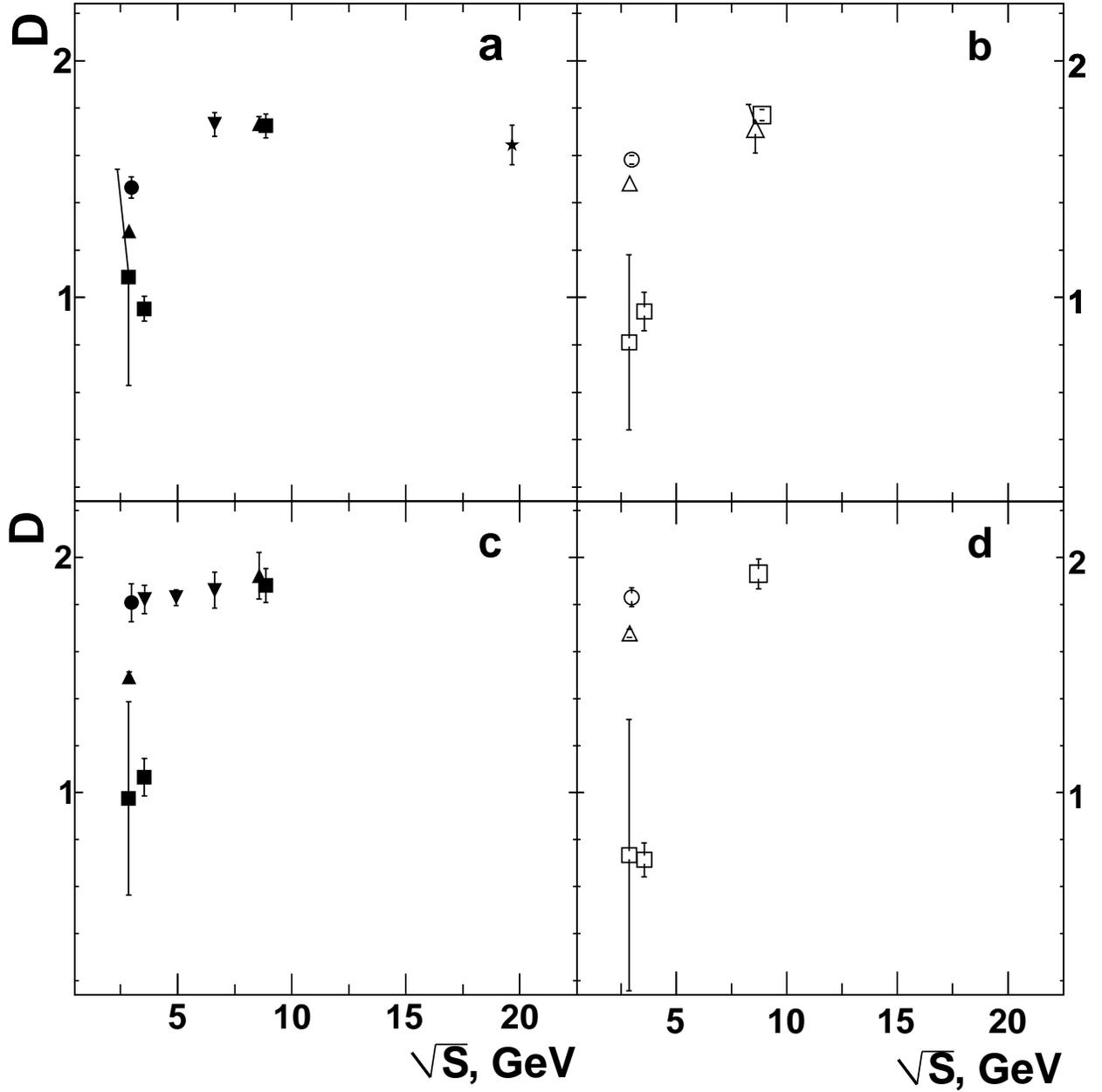}
\caption{Dependence of the cluster dimension $D$,
calculated for $\pi^{-}$-mesons, on collision energy at
$X_{\footnotesize \mbox{l}}=0.1$ (a,b) and $X_{\footnotesize
\mbox{l}}=0.2$ (c,d). Left hand corresponds to the target
fragmentation region (a,b), right hand -- to the beam
fragmentation (b,d). Experimental points for target (beam)
fragmentation region are marked as follows: {\large$\bullet$}
({\large$\circ$}) -- $\pi^{+}\mbox{p}$ at 4.2 GeV/c,
$\blacktriangledown$ -- $\bar{\mbox{p}}\mbox{p}$ at 5.7, 12 and
22.4 GeV/c, {\large$\star$} -- $\mbox{pp}$ at 205 GeV/c,
$\blacktriangle$ ($\vartriangle$) -- $\pi^{-}\mbox{p}$ at 3.9 and
40 GeV/c, $\blacksquare$ ($\square$) --
$\pi^{-}\left(\mbox{C}_{2}\mbox{F}_{5}\mbox{Cl}_{3}\right),~\pi^{-}\mbox{Ne},~\pi^{-}\mbox{C}$
at 3.9, 6.2 and 40 GeV/c,
respectively.}\label{fig:4-ClusterDim-vs-Energy}
\end{figure}

Fig.\ref {fig:4-ClusterDim-vs-Energy} shows cluster dimension $D$,
calculated for secondary $\pi^{-}$-mesons, versus collision energy
$\sqrt{s}$ for hadron-hadron and hadron-nuclear interactions at
$X_{\footnotesize \mbox{l}}=0.1$ (a,b) and at
$X_{\footnotesize\mbox{l}}=0.2$ (c,d) for region of target
fragmentation (a,c) and for beam fragmentation (b,d). Experimental
data used for calculation of cluster dimension are taken from
\cite{Baldin-PrJINR-142-1987} for $\bar{\mbox{p}}\mbox {p}$ and
$\mbox{pp}$ interactions, from
\cite{Baldin-PrJINR-142-1987,Baldin-ECHAYA-29-577-1998} -- for
$\pi^{-}\mbox{p}$ and $\pi^{-}\mbox{C}$ at 40 GeV/c. Cluster
dimension shows fractional value for all reactions under study
with exception of hadron-nuclear interactions (\ref {eq:reac-4})
and (\ref {eq:reac-5}), that allows to suggest the presence of
indication on manifestation of fractal-like properties of pion
jets in collision energy range under considered. The parameter $D$
have values close to the integer ones within error bars for
reactions (\ref {eq:reac-4}) and (\ref {eq:reac-5}), excepting for
$\pi^{-}\mbox{Ne}$ reaction for beam fragmentation at more hard
cutoff $X_{\footnotesize\mbox{l}}=0.2$. Accessible experimental
data allow to study the behaviour of $D\left(\sqrt {s}\right)$ and
some features of the dependence only at a qualitative level. At
any used $X_{\footnotesize\mbox{l}}$ cluster dimension for
reactions (\ref {eq:reac-4}) and (\ref {eq:reac-5}) is close to
the integer value within larger errors as indicated above, results
obtained for hadron-nuclear reactions show increasing of $D$ with
growth of initial energy. The hadron-hadron interactions show a
different behaviour of $D\left(\sqrt {s}\right)$ in various range
of initial energies for region of target fragmentation.
Significant growth of cluster dimension is observed at small
increasing of collision energy for reactions (\ref {eq:reac-1})
and (\ref {eq:reac-3}), the further increasing of $\sqrt{s}$ leads
to much weaker growth of $D$ with the subsequent reaching of a
constant at $X_{\footnotesize\mbox{l}}=0.1$
(Fig.\ref{fig:4-ClusterDim-vs-Energy}a) or to absence of changes
at all for $X_{\footnotesize\mbox{l}}=0.2$
(Fig.\ref{fig:4-ClusterDim-vs-Energy}c). The dependencies
$D\left(\sqrt{s}\right)$ for beam fragmentation region
(Fig.\ref{fig:4-ClusterDim-vs-Energy}b,d) do not contradict with
behaviour of corresponding dependencies described above for a
target fragmentation. The observed behaviour of cluster dimension
versus collision energy and sharp amplification of manifestation
of fractal-like properties, which is one of the important and
characteristic features for occurrence of new degrees of freedom,
does not contradict a hypothesis about presence of transition
energy range to experimentally noticeable manifestation of quark
degrees of freedom in hadron jet production at $\sqrt{s} \sim 3$
GeV.

There is an intensive production of meson resonances in the energy
range under study. Influence of resonances on cluster dimension of
pion jets has been investigated for following exclusive channels
$\pi^{-}+\mbox{p} \to \mbox{p}+\pi^{+}+2\pi^{-},~\pi^{+}+\mbox{p}
\to \mbox{p}+2\pi^{+}+\pi^{-}$ of reactions (\ref {eq:reac-1}) and
(\ref {eq:reac-3}), respectively. For jets (clusters) of particles
with identical masses there is the unambiguous relation between
$\langle b_{k} \rangle$ and jet effective mass
\cite{Okorokov-IntSemHEP-154-2006,Grishin-Book-1988}. In
consequence of this relation meson resonances, in particular,
clear identified $\rho(700)$ and $a_{1}(1320)$, give rise to the
normalized dependence of number of particles in jet on its radius,
$N(b_{k})/N_{\footnotesize \mbox {tot}}$, in value range $3 \leq
b_{k} < 4$ for beam fragmentation region. The influence of the
specified resonances on cluster dimension is negligible for
$\pi^{-}\mbox{p}$ exclusive channel under study at soft condition
of selection $X_{\footnotesize\mbox{l}}=0.1$ and significantly
larger at hard cutoff. But the influence of these resonances is
observable for corresponding $\pi^{+}\mbox{p}$ channel at any
$X_{\footnotesize\mbox{l}}$. The contribution from decay of meson
resonances leads to significant growth of fractal cluster
dimension of pion jets. Influence of resonances has been
investigated in details for exclusive channels with one $\pi^{0}$
in reactions (\ref {eq:reac-1}) and (\ref {eq:reac-3}) also.
However because of narrowness of $\eta$ and $\omega$ mesons it is
possible to speak only about presence of indication on the
influence of these resonances on cluster dimension.

\section{\label{sec4}Summary}
In conclusion, we summarize the main results of this study.

Values of $\langle b_{k}\rangle$ are significantly smaller for
reactions (\ref {eq:reac-1}) - (\ref {eq:reac-3}) at $\sqrt{s}
\sim 3$ GeV than those in other interactions at slightly higher
energies, that leads to changing of dependence $ \langle
b_{k}\rangle \left(\sqrt{s}\right)$ from logarithmic behaviour to
power one. It is possible to assume, that this effect is caused by
occurrence of experimentally noticeable manifestation of quark
degrees of freedom in soft pion jet production and corresponding
transition from the description of the such process at language of
nucleon-meson degrees of freedom to using of color (quark-gluon)
degrees of freedom. The behaviour of dependence $\langle
T_{k}\rangle \left(\sqrt{s}\right)$ for hadron-hadron interactions
qualitatively confirms this hypothesis.

The cluster dimension of soft pion jets has been obtained for
various reactions at $\sqrt{s} \sim 3-20$ GeV for the first time.
The dimension shows fractional value for main part of reactions
under considered that allows to suggest the presence of
fractal-like properties for pion jets. Features of behaviour of
dependence $D \left(\sqrt{s}\right)$ obtained for hadron-hadron
reactions do not contradict the above suggestion of manifestation
of new degrees of freedom at $ \sqrt{s} \sim 3$ GeV. Influence of
meson resonances is observed on fractal dimensions of pion jets in
beam fragmentation region, leading to significant increasing of
$D$.


%
\newpage
\newpage
\newpage
\newpage
\end{document}